\documentstyle[12pt]{article}
\newlength{\dinwidth}
\newlength{\dinmargin}
\newcommand{\ba}{\begin{array}}
\newcommand{\ea}{\end{array}}
\newcommand{\be}{\begin{equation}}
\newcommand{\ee}{\end{equation}}
\newcommand{\bea}{\begin{eqnarray}}
\newcommand{\eea}{\end{eqnarray}}

\def\bee{\begin{eqnarray}}
\def\eee{\end{eqnarray}}
\def\be{\begin{equation}}
\def\ee{\end{equation}}

\newcommand{\beas}{\begin{eqnarray*}}
\newcommand{\eeas}{\end{eqnarray*}}

\font\cmss = cmss12

\def\integer{{\rlap{\cmss Z} \hskip 1.8pt \hbox{\cmss Z}}}
\def\laplace{{\kern1pt\vbox{\hrule height 1.2pt\hbox{\vrule width 1.2pt\hskip
  3pt\vbox{\vskip 6pt}\hskip 3pt\vrule width 0.6pt}\hrule height 0.6pt}
  \kern1pt}}
\def\scriptlap{{\kern1pt\vbox{\hrule height 0.8pt\hbox{\vrule width 0.8pt
  \hskip2pt\vbox{\vskip 4pt}\hskip 2pt\vrule width 0.4pt}\hrule height 0.4pt}
  \kern1pt}}

\def\roughly#1{\raise.3ex\hbox{$#1$\kern-.75em\lower1ex\hbox{$\sim$}}}

\begin{document}
\thispagestyle{empty}
\addtocounter{page}{-1}
\begin{flushright}
SNUTP 97-152\\
{\tt hep-th/9711081}\\
\end{flushright}
\vspace*{1.3cm}
\centerline{\Large \bf Gravitating M(atrix) Q-Balls
\footnote{
Work supported in part by the NSF-KOSEF Bilateral Grant, 
KOSEF SRC-Program, Ministry of Education Grant BSRI 97-2410, 
and the Korea Foundation for Advanced Studies.}}
\vspace*{1.2cm} \centerline{\large\bf Soo-Jong Rey}
\vspace*{0.8cm}
\centerline{\large\it Physics Department, Seoul National University,
Seoul 151-742 KOREA}
\vspace*{0.6cm}
\centerline{\large\tt sjrey@gravity.snu.ac.kr}
\vspace*{1.5cm}
\centerline{\large\bf abstract}
\vskip0.5cm
Q-ball configuration that represents oscillating or spinning closed
membrane is constructed via M(atrix) theory. Upon gravitational collapse
Q-balls are expected to form Schwarzschild black holes. For quasi-static
spherical membrane, we probe spacetime geometry induced by monopole
moment via D0-parton scattering off the Q-ball. We find a complete
agreement with long distance potential calculated using eleven-dimensional
supergravity. Generalizing to heterotic M(atrix) theory, we also construct
Q-ball configurations of real projective and disk membranes. The latter
Q-ball configuration arises as twisted sector of heterotic M(atrix) theory,
hence, are expected to form a charged black hole after 
gravitational collapse.
\vspace*{1.1cm}

%\centerline{Submitted to Nuclear Physics B}
\setlength{\baselineskip}{18pt}
\setlength{\parskip}{12pt}

\newpage
%%%%%%%%%%%%%%%%%%%%%%%%%%%%%%%%%%%%%%%%%%%%%%%%%%%%%%%%%%%%%%%%%%%%%%%%
\section{Introduction}
%%%%%%%%%%%%%%%%%%%%%%%%%%%%%%%%%%%%%%%%%%%%%%%%%%%%%%%%%%%%%%%%%%%%%%%%
Consider clusters of D0-partons in M(atrix) theory~\cite{bfss}, the 
light-front 
description of M-theory~\cite{witten}. If a macroscopically large number of 
D0-partons are clustered together, one expects that they collapse
under the influence of gravity. Eventually one will see that the D0-partons
form a black hole. From Type IIA string theory point of view, this is the
formation of black hole out of D-branes at strong coupling limit. As such,
M(atrix) theory offers an exciting possibility of understanding quantum aspects
of a black hole directly in the black hole regime~\cite{dvv, martinec, mine}.
Indeed, it has been argued recently~\cite{susskind, klebanovsusskind, horowitzmartinec, susskind2} that the entropy of Schwarzschild black holes can
be derived from M(atrix) theory at near extreme limit, at least, at special
situation where the number of D0-partons $N$ is equal to the entropy $S$.
Eventually, utilizing M(atrix) theory formulation, one would like to understand
{\sl dynamics} of quantum black holes such as formation and Hawking radiation.

One may consider, in M(atrix) theory, an alternative possibility of black hole
formation. Instead of clustering into threshold bound-states,
D0-partons may arrange themselves into Landau-orbiting bound-state of an 
incompressible two-dimensional fluid and form a membrane first.
In non-compact spacetime, such a membrane is unstable
and will collapse via tension and gravitational attraction.
Eventually it will form a black hole once the size shrinks smaller
than its Schwarzschild radius. As a first step to investigation of this
extremely interesting process, in this paper, we study collapsing or
oscillating membrane dynamics in M(atrix) theory. Because its proximity
to non-topological solitons~\cite{coleman}, we call such membrane configuration
as M(atrix) Q-ball~\footnote{While this work was in final stage, we have
received a preprint by Horowitz and Martinec~\cite{horowitzmartinec} in which section
4 discusses related issues.}.

In section 2, we construct an exact solution of spherical M(atrix) Q-ball.
M(atrix) membrane of spherical topology has already been
constructed~\cite{kimrey}, based on earlier results~\cite{dewit}.
We extend the result by studying kinematics and, in particular,
{\sl dynamics} of such membrane. 
The spherical M(atrix) membrane is closely related to monopoles of
magnetic charge 2 in $N=4$ supersymmetric Yang-Mills theory.
Utilizing previous investigation, we derive an exact classical solution
of the spherical M(atrix) membrane.

Curved spacetime geometry will be induced around a spherical membrane.
One possible M(atrix) theory probe of the geometry is to scatter D0-partons 
off the spherical M(atrix) membrane. We have calculated `monopole' part of
the potential. The calculation is tedious but straightforward and 
turns out to agree with result deduced from classical supergravity.

Quasi-stable M(atrix) membranes of other topologies are also possible.
The simplest ones are disk and real-projective membranes. They are 
appropriate $\integer_2$ involution configuration of the spherical membrane.
They arise naturally in heterotic M(atrix) theory in which the $\integer_2$
projection provides an $\Omega9$ orientifold plane and end-of-world nine branes.
Spherical membrane stuck at the orientifold plane or the end-of-world nine
brane is nothing but the membranes of real-projective and disk 
topology. Construction of these membranes in heterotic M(atrix) theory is
also given~\cite{kimrey}. In this paper, we construct disk and real-projective
M(atrix) membranes explicitly and study their kinematics and dynamics. 
Since they are associated with heterotic M(atrix) theory with
only eight superchrages, the induced spacetime geometry is expected to
exhibit far richer physics.

Our notation of M(atrix) theory is as follows.
Regularizing zero-momentum limit by compactifying the
longitudinal direction on a circle of radius $R$, the M(atrix) theory
action is given by a matrix quantum mechanics of $SU(N)$ gauge group
\cite{bfss} :
\be
S_M = {\rm Tr}_N \int \! d\tau \, \Big(
{1 \over 2 R} (D_\tau {\bf X}^I)^2 + {R \over 4} [{\bf X}^I , {\bf X}^J]^2
+ {\bf \Theta}^T D_\tau {\bf \Theta} + i R {\bf \Theta}^T \Gamma_I [{\bf X}^I ,
{\bf \Theta}] \Big).
\label{action}
\ee
Here, ${\bf X}^I$ and ${\bf \Theta}^\alpha$ denote 9 bosonic and 16 spinor
coordinates of 0-brane partons ($I = 1, \cdots, 9$ and $\alpha = 1, \cdots,
16$). The Majorana spinor conventions are such that $\Gamma_I$'s are real
and
symmetric and $i {\overline {\bf \Theta}} \Gamma_- \equiv {\bf \Theta}^T$:
\be
\Gamma_i = \left( \begin{array}{cc}
0 & \sigma^i_{a {\dot a}} \\ \sigma^i_{{\dot a} a} & 0 \end{array} \right)
\hskip0.5cm i = 1, \cdots, 8; \hskip1cm
\Gamma_9 = \left( \begin{array}{cc}
- \delta_{a b} & 0 \\ 0 & + \delta_{{\dot a} {\dot b}} \end{array} \right).
\label{gammamatrix}
\ee
The non-dynamical gauge field $A_\tau$ that enters through covariant
derivatives $D_\tau {\bf X}^I \equiv \partial_\tau {\bf X}^I -
i [A_\tau, {\bf X}^I]$ and
$D_\tau {\bf  \Theta}^\alpha \equiv \partial_\tau {\bf \Theta}^\alpha -
i [A_\tau, {\bf \Theta}^\alpha]$ projects the physical Hilbert space to
a gauge singlet sector and ensures invariance under area-preserving
diffeomorphism transformation.
Hamiltonian in the infinite momentum limit is given by
\be
H_M = R \, {\rm Tr}_N \Big( {1 \over 2} {\bf \Pi}^2_I
- {1 \over 4} [{\bf X}^I, {\bf X}^J]^2 + i {\bf \Theta}^T \Gamma_I [{\bf X}^I,
{\bf \Theta}] \Big).
\label{hamiltonian}
\ee
The M(atrix) theory is invariant under the following supersymmetry
transformations
\bee
\delta {\bf X}^I &=& - 2 \epsilon^T \Gamma^I {\bf \Theta} \nonumber \\
\delta {\bf \Theta} &=& { i \over 2} \Big( \Gamma_I D_\tau {\bf X}^I
+ {1 \over 2} \Gamma_{IJ} [{\bf X}^I, {\bf X}^J] \Big) \, \epsilon + \xi
\nonumber \\
\delta A_\tau &=& - 2 \epsilon^T \, {\bf \Theta}.
\label{susytransformation}
\eee
Here, $i \epsilon$ is a 16-component spinor generator of local
supersymmetry, while $\xi$ is a 16-component spinor generator of rigid
translation.
As such, the sixteen dynamical and sixteen kinematical supersymmetry charges
are given by:
\bee
{\bf Q}_\alpha &=& {\sqrt R} {\rm Tr} \Big( \Gamma^I {\bf \Pi}_I +
{i  \over 2} \Gamma_{IJ} [{\bf X}^I, {\bf X}^J] \Big)_{\alpha \beta}
{\bf \Theta}_\beta,
\nonumber \\
{\bf S}_\alpha &=& {2 \over \sqrt R} {\rm Tr} {\bf \Theta}_\alpha
\label{susycharges}
\eee
respectively.

This paper is organized as follows. In section 2, we study kinematics
and dynamics of macroscopically large spherical M(atrix) Q-ball.
The configuration will induce spatial curvature around it. In
section 3, we probe the geometry by scattering off the D0-parton and 
find complete agreement with results derived from classical supergravity. 
In section 4, we extend the result of section 2 to macroscopic M(atrix)
Q-ball of disk and real-projective topologies. These configurations are 
relevant for heterotic M(atrix) theory, in which the disk and real-projective
M(atrix) Q-balls arise as quasi-static configurations of `twisted sector' 
at end-of-world nine brane or orientifold. Discussions are relegated in 
section 5.

%%%%%%%%%%%%%%%%%%%%%%%%%%%%%%%%%%%%%%%%%%%%%%%%%%%%%%%%%%%%%%%%%%%%%%%
\section {Spherical M(atrix) Q-Ball}
%%%%%%%%%%%%%%%%%%%%%%%%%%%%%%%%%%%%%%%%%%%%%%%%%%%%%%%%%%%%%%%%%%%%%%%
Let us begin with construction of a spherical M(atrix) Q-ball. 
Essential observation is a decade old~\cite{dewit}, and is based on
the fact that a spherical Q-ball should manifest area-preserving 
diffeomorphism Diff($S_2)$, which is residual gauge symmetry of quantized
membrane in light-cone gauge.

We follow the presentation in Ref.~\cite{kimrey} and describe the 
area-preserving diffeomorphism of Diff($S_2$) .
Introduce a complete set of scalar spherical harmonics
\be
Y_{\bf lm} ({\bf x}) \equiv
C_{i_1, \cdots, i_l} x^{i_1} \cdots x^{i_l}
\label{sphericalharmonics}
\ee
in terms of embedding space coordinates ${\bf x} \equiv
(x_1, x_2, x_3)$ satisfying
\be
{\bf x} \cdot {\bf x} = 1.
\ee
In Eq.(\ref{sphericalharmonics}), $C_{i_1, \cdots, i_n}$ are symmetric,
traceless, tensor coefficients. The `magnetic' quantum number
ranges over $\bf m$: $ - {\bf l} \le {\bf m} \le  {\bf l}$.
The area-preserving diffeomorphism algebra Diff($S_2$) is encoded via the 
Poisson bracket algebra among the spherical harmonics
\bee
\{ Y_{\bf lm}, , Y_{{\bf l'm'}} \} & \equiv & \epsilon_{ijk} x^i
(\partial_j Y_{\bf lm}) ( \partial_k Y_{{\bf l'm'}})
\nonumber \\
&=& \bigoplus_{{\bf j} = |{\bf l} - {\bf l}'| + 1}^{{\bf l} + {\bf l}' - 1}
[\, Y_{\bf j(m+m')} \, ],
\label{bracket}
\eee
viz. a sum of irreducible polynomials of scalar harmonics in the range
$|{\bf l} - {\bf l}' | + 1 \le {\bf j} \le ({\bf l} + {\bf l}' - 1)$.

Important observation of Ref.~\cite{dewit} was that the above construction 
of area-preserving diffeomorphism algebra is in one-to-one
correspondence with $SU(N)$ Lie algebra expressed in terms of maximal
embedding of $SU(2)$. Under maximal embedding,
the generators of $SU(N)$ can be expressed as products of $SU(2)$
generators $\Sigma_i$ in the $N$-dimensional representation:
\bee
T^{(1)} &=& C_i \Sigma_i
\nonumber \\
T^{(2)} &=& C_{ij} \Sigma_i \Sigma_j \nonumber \\
\cdots
\nonumber \\
T^{(N-1)} &=& C_{i_1 i_2 \cdots i_{N-1}} \Sigma_{i_1} \cdots
\Sigma_{i_{N-1}}.
\label{sungenerators}
\eee
Here, the coefficients $C_{ijk \cdots}$ are the same symmetric, traceless,
tensor coefficients as in Eq.(\ref{sphericalharmonics}).
The above form of $SU(N)$ generators expresses simply that,
using the fact that the fundamental $N$-dimensional
representation of $SU(N)$ remains irreducible in $SU(2)$, the adjoint
representation of $SU(N)$ is decomposed into $N^2 - 1 =
3 + 5 + \cdots + (2N-1)$ representations of $SU(2)$.
The $T^{(i)}$ matrices form a complete set of traceless, Hermitian
$N \times N$ matrices, hence, provide a basis for $SU(N)$.

The $\Sigma_i$ generators in the $N$-dimensional representation of $SU(2)$
can be represented by a totally symmetrized $2^{(N-1)} \times
2^{(N-1)}$ matrices:
\be
\Sigma_i = {\rm Sym} [
\sigma_i \otimes {\bf I} \otimes \cdots \otimes {\bf I}+
{\bf I} \otimes \sigma_i \otimes {\bf I} \otimes \cdots \otimes {\bf I}
+\cdots+{\bf I} \otimes {\bf I} \otimes \cdots \otimes \sigma_i], \hskip0.5cm
i=1,2,3
\label{su2generators}
\ee
in which $\sigma_i, i = 1,2,3$ are the Pauli matrices.

Comparison of Eq.(\ref{sphericalharmonics}) and Eq.(\ref{sungenerators})
shows that the commutation relation among $SU(N)$ generators $T_{(i)}$
is isomorphic to the Poisson bracket relation Eq.(\ref{bracket}) among the
spherical harmonics $Y_{\bf l}({\bf x})$. In the large $N$ limit,
the non-commutativity among $T^{(i)}$'s becomes irrelevant. Thus,
the area-preserving diffeomorphism of sphere ${\rm Diff}_0 (S_2)$ is
realized by $SU(N)$ algebra, reproducing the well-known result\cite{dewit}.

Accordingly, the M(atrix) fields ${\bf X}^I$ representing transverse 
coordinates of the spherical membrane can be expanded in terms of the 
SU(2) generators
\be
{\bf X}^I(t) = \sum_{n=1}^{N-1} \hbar^n
 C_{i_1 i_2 \cdots i_{n-1}}^I (t) \Sigma_{i_1} \cdots \Sigma_{i_{n-1}}
\label{expansion}
\ee
where we have introduced a notation $\hbar \equiv 2 \pi / N$. 
In terms of the expansion, we now look for Q-ball solution. In
$A_0 = 0$ gauge, M(atrix) equations of motion
read
\be
\partial_\tau^2  {\bf X}^I + R^2  [{\bf X}^J, [{\bf X}^J, {\bf X}^I]] = 0.
\ee
supplemented with the Gauss' law constraint:
\be
[{\bf X}^I, \partial_\tau {\bf X}^I ] = 0.
\ee

An exact solution of the above M(atrix) equation of motion 
in the form of  Eq.~(\ref{expansion}) is generally not possible. This is 
because the equation is nonlinear and mode-mode coupling requires to solve
all the higher harmonics generated via Eq.~(\ref{bracket}). There is, however,
 one 
and unique exception to this. From Eq.~(\ref{bracket}), we observe that, if
${\bf l} = {\bf l'} = 1$, then the harmonics produced again has ${\bf j} = 1$.
Thus, a unique exact solution of spherical M(atrix) Q-ball is given by
\be
{\bf X}^i (\tau) = \hbar X(t) \cdot \Sigma^i
\quad \quad\quad \quad \quad (i = 1,2,3)
\ee
where $R(t)$ is a time-dependent function representing `breathing' 
excitation of the Q-ball. 
In this case, the SU(2) generators $\Sigma^I$ are in one-to-one correspondence
with embedding cartesian coordinates 
\be
(\Sigma^1, \Sigma^2, \Sigma^3) \quad \quad \longleftrightarrow 
\quad \quad
(\cos \theta, \sin \theta \cos \phi, \sin \theta \sin \phi)
\ee
where $(\theta, \phi), \quad \theta = [0, \pi], \, \phi=[0, 2\pi]$ 
is the spherical coordinates of $S_2$.
One can also take the breathing function anisotropically $X(\tau) \rightarrow 
X^i (\tau)$. The 
corresponding M(atrix) Q-ball represent a close membrane whose shape is
ellipsoidal~\footnote{Closely related configurations have been studied
previously in the context of supermembrane~\cite{bergshoeff}. See also the
second and third references of Ref.~\cite{dewit}.}
Indeed, the second Casimir invariant
\be
{\rm Tr} {\bf X} \cdot {\bf X} = \hbar^2 \sum_{i=1,2,3} (X^i(\tau))^2
C_2 
\quad \quad \quad \quad \quad 
C_2 \equiv {\rm Tr} \Sigma_i \Sigma_i = (N^2 - 1)/4
\ee
can be interpreted as repsentation of the ellipsoid.

The equation of motion for $X(\tau)$  
\be
\partial_\tau^2 X(\tau) + R^2 X^3(\tau) = 0
\ee
is solved by the first integral
\be
{1 \over 2} ( \partial_\tau X)^2 + {R^2  \over 4} X^4 (\tau) 
 = E.
\ee
We have assumed that the Q-ball has nonzero conserved energy $E$. The 
Q-ball configuration represents a breathing membrane of spherical topology
. The spherically symmetric Q-ball is special in that the three major axis
are retained throughout the evolution. On the other hand, squeezed 
(ellipsoidal) membranes will deform in $X^-$ direction as is evident from
the integral of light-cone Hamiltonian. 

The above spherical Q-ball is not stable. 
Initially, even though arranged initially at rest with a size $(E/R^2)^{1/4}$,
the Q-ball will eventually shrink to a point and expand to the 
opposite direction with reversed orientation. The Q-ball continues 
the vibrational motion. As the Q-ball exhibits a highly anisotropic transverse 
motion and is also electrically charged, it will eventually dissipate away
its energy via radiation of M-theory graviton and three-form tensor potential. 

Actually, it is possible to stabilize the spherical Q-ball against collapse.
If we spin up the Q-ball, centrifugal repulsion can balance the Q-ball against 
collapse. Such a spinning Q-ball is given by~\footnote{Hoppe~\cite{hoppe} has 
originally studied rotating membrane configuration, but in different topology.
Another closely related configurations have been studied recently 
in~\cite{fairlie}.}
\be
{\bf X}^i (\tau) = \hbar X(\tau) \cdot \Omega^{ij} (t) \cdot \Sigma^j
\quad \quad \quad \quad \quad (i=1,2,3)
\ee
where
$\Omega (t) \equiv \exp ( \phi(\tau) {\bf T})$ represents rotational excitation. The matrix ${\bf T}$ is a constant
SU(2) matrix such that ${\bf T}^2 \Sigma = - \mu \Sigma$. 
The angular momentum is conserved:
 \be
X^2(\tau) (\partial_\tau \phi (\tau)) \equiv L = {\rm constant}
\ee
and the first integral of equation of motion is given by
\be
{1 \over 2} (\partial_\tau X)^2 + {R^2 \over 4}
X^4 + { \mu L^2 \over 2} { 1 \over X^2} = E
\ee
It is clear that the spinning Q-ball is stabilized
at breathing radius $X = \left( \mu L^2 / R^2 \right)^{1/5}$.
Nevertheless, the spinning Q-ball is highly anisotropic
in nine-dimensional transverse space. As such, it is expected that
the spinning Q-ball also collapses gradually as it radiates away graviton 
and antisymmetric fields. 

It is to be noted that a minimal Q-ball can be constructed out of SU(2) 
M(atrix) gauge group, viz. two D0-partons. These miniature Q-balls may be
viewed as fundamental building unit of a generic macroscopic Q-ball. 
Spherical Q-balls will interact each other and can change 
topologies, and we may view  macroscopic spherical Q-ball made out of order
${\cal O}( N/2)$ miniature Q-balls.

%%%%%%%%%%%%%%%%%%%%%%%%%%%%%%%%%%%%%%%%%%%%%%%%%%%%%%%%%%%%%%%%%%%%%%%
\section{Spacetime Geometry around Spherical Membrane}
%%%%%%%%%%%%%%%%%%%%%%%%%%%%%%%%%%%%%%%%%%%%%%%%%%%%%%%%%%%%%%%%%%%%%%%
The spherical M(atrix) Q-ball is gravitating and will form a black hole
eventually. It is therefore of interest to derive the induced spacetime
geometry directly from M(atrix) theory itself. In this section, via
D0-parton scattering, we probe the spacetime geometry. Using the background
field method, we calculate the velocity-dependent potential between the
D0-parton and the spherical Q-ball. As we will see momentarily, the result
turns out to be in complete agreement with the calculation based on eleven-dimensional
supergravity.

%%%%%%%%%%%%%%%%%%%%%%%%%%%%%%%%%%%%%%%%%%%%%%%%%%%%%%%%%%%%%%%%%%%%%%%%%%
\subsection{M(atrix) Theory Calculation}
%%%%%%%%%%%%%%%%%%%%%%%%%%%%%%%%%%%%%%%%%%%%%%%%%%%%%%%%%%%%%%%%%%%%%%%%%%
Let us first evaluate the long-range potential between the D0-parton and
the non-rotating spherical Q-ball. 
For a general background M(atrix) theory configuration, general expression of 
one-loop effective action has been found and we gratefully make use of the 
result~\cite{tseytlin}.
Adapting to the present case, where D0-parton is used as a probe, the leading
infrared correction at one-loop ${\bf \Gamma}^{(1)}$ is given by
\be
{\bf \Gamma}^{(1)}[r, F]
= - { \Gamma(7/2) \over 2 (4 \pi)^{1/2} r^7}
\int d \tau \, V^{(1)}
\ee
where
\be
V^{(1)} = {2 \over 3} {\rm Tr}
\left( F_{ab} F_{bc} F_{cd} F_{da} + {1 \over 2}
F_{ab} F_{bc} F_{da} F_{cd} - {1 \over 4} F_{ab} F_{ab} F_{cd} F_{cd}
-{1 \over 8} F_{ab} F_{cd} F_{ab} F_{cd} \right).
\ee
Here, Tr denotes trace in the adjoint representation and $F_{ab} =
({\bf E}_i, {\bf B}_i)$ are the background gauge field configuration.
The formula is most conveniently derived using background field methos
in covariant gauge~\cite{lifschytz}. 
Keeping only the off-diagonal excitations between the D0-parton and the
Q-ball configuration, the above formula is derived from a graded sum over 
harmonic oscillator frequencies
for ten bosonic fields, two bosonic ghosts, and sixteen fermionic fields
(all counted in complex unit).

Inserting the Q-ball:
\be
{\bf E}_i \equiv (\partial_\tau {\bf X}_i) = \hbar \dot X \cdot \Sigma_i
\quad \quad \quad \quad
{\bf B}_i \equiv {1 \over 2} \epsilon_{ijk} [{\bf X}_j, {\bf X}_k] = 
\hbar^2 X^2 \cdot \Sigma_i
\ee
and a single D0-parton configuration, 
we find that the M(atrix) interaction potential is provided by the 
`self-energy' of the M(atrix) Q-ball  and is given by
\be
V(r) = - {15 \over 8} {1 \over r^7} \hbar R^2 M^2 
\left( 1 + {\cal O}({1 \over N^2}) \right),
\label{matrix}
\ee
where $M$ denotes the self-energy of M(atrix) Q-ball
\bee
M &=& {1 \over R} {\rm Tr} \Big[
{1 \over 2} (\partial_\tau {\bf X}^i)^2 - {1 \over 4} [{\bf X}^i, {\bf X}^j]^2
\Big]
\nonumber \\
&=& {1 \over R} {\rm Tr} (\Sigma_i \Sigma_i) 
\cdot \Big[ {1 \over 2} \hbar^2 (\dot X)^2 .
+ {\mu L^2 \over 2} \hbar^2 {1 \over X^2} + {1 \over 4} \hbar^4 X^4 \Big].
\eee
Note that the large $N$ correction begins to appear only at order 
${\cal O}(1/N^2)$.

%%%%%%%%%%%%%%%%%%%%%%%%%%%%%%%%%%%%%%%%%%%%%%%%%%%%%%%%%%%%%%%%%%%%%%%%%%
\subsection{Supergravity Calcuation}
%%%%%%%%%%%%%%%%%%%%%%%%%%%%%%%%%%%%%%%%%%%%%%%%%%%%%%%%%%%%%%%%%%%%%%%%%%
In supergravity, the one-graviton exchange between the longitudinal
graviton and the boosted Q-ball is the leading order effect.
The energy-momentum tensor of the longitudinal graviton representing
a single D0-partons is given by $p^+ = 1/R$ viz.
\be
T_{--} = {1 \over R} \delta (x^-).
\ee
For Q-ball configuration, we take an adiabatic approximation under the 
assumption that the configuration is quasi-static. Then, the Q-ball 
configuration is described by an energy-momentum source of `monopole'
component with longitudinal momentum $p^+ = N/R$ and light-cone energy $E = M$.
We also assume that the D0-parton propagates at a large impact parameter $r$ 
from the center of the Q-ball configuration. 

In the light-front coordinates, it is straightforward to evaluate the 
one-graviton exchange between the massless eleven-dimensional particle 
representing the D0-parton  and the light-front monopole component of the
boosted Q-ball. The potential is found to be 
\be
V_{\rm sugra} = - {15 \over 8} \hbar {1 \over r^7} R^2 M^2,
\label{sugra}
\ee
where we have ignored possible radiation and back-reaction effects.
At least for initial stage of the collapse and low angular velocity, 
this should be a good approximation.
The supergravity potential Eq.~(\ref{sugra}) is precisely the same result 
as the M(atrix) theory potential calculated in Eq.~(\ref{matrix}).

To summarize, we have found exact agreement of the interaction potential
between Q-ball and D0-parton as calculated both in M(atrix) theory and
in supergravity up to ${\cal O}(1/N^2)$ corrections.

%%%%%%%%%%%%%%%%%%%%%%%%%%%%%%%%%%%%%%%%%%%%%%%%%%%%%%%%%%%%%%%%%%%%%%%
\section{Heterotic M(atrix) Q-Balls}
%%%%%%%%%%%%%%%%%%%%%%%%%%%%%%%%%%%%%%%%%%%%%%%%%%%%%%%%%%%%%%%%%%%%%%%
So far we have considered M(atrix) Q-balls of spherical shape. It is 
also possible to extend the investigation to quasi-stable M(atrix) Q-balls. 
of other topology. Of particular interest are the ones with disk or 
real-projective topologies. 
While they do not arise in M(atrix) theory with sixteen supercharges, 
disk and real-projective M(atrix) Q-balls are possible in heterotic
M(atrix) theory~\cite{kimrey, rey} or non-orientable M(atrix) theories~\cite{
kimrey2}, where twisted sector provides end-of-world nine-branes
and orientifolds where the disk and real-projective M(atrix) Q-balls 
can appear.

Let us begin with disk M(atrix) Q-ball, which may arise as a membrane
configuration attached to the end-of-nine brane in heterotic M(atrix)
theory. Most straightforwardly disk Q-ball can be described by an involution
\be
\integer_2 \quad : \quad  (\theta, \phi) \quad \longleftrightarrow 
(\pi - \theta, \phi + \pi),
\label{involution}
\ee
from the spherical Q-ball, under which
\be
Y_{\bf lm} ({\bf x}) \rightarrow (-)^{\bf l + m} Y_{\bf l m} ({\bf x}).
\label{polarinvolution}
\ee
The vector harmonics that form a basis of generators of
area-preserving diffeomorphism of disk Q-ball are obtained as 
parity-odd combinations:
\be
L_{\bf lm} \equiv \{ Y_{\bf lm}: Y_{\bf lm} - (-)^{\bf l+m} Y_{\bf lm} \}.
\label{d2subset}
\ee
Since only odd values of $\bf
(l+m)$ are selected as the basis, the total number
of vector harmonics generators are given by
\be
L_{\bf lm} = \{ Y_{1,0}, \, Y_{2,+1}, \, Y_{2,-1}, \, Y_{3,+2} , \, Y_{3,0},
\, Y_{3, -2} , \cdots \},
\ee
hence, yield $1 + 2 + 3 + \cdots + (2N - 1) = 2 N ( 2N - 1) / 2$ generators.
This equals precisely to the number of generators of $SO(2N)$ group, 
and the area-preserving diffeomorphism of disk is described by $N
\rightarrow \infty$ limit of $SO(2N)$ Lie algebra \footnote{
A similar analysis shows that $SO(2N+1)$ subgroup is also possible by recalling
that $1 + 2 + \cdots + (N-1) = N(N-1)/2$. }.

It can also be shown that the above construction of disk area-preserving
diffeomorphism is in one-to-one correspondence with the Lie algebra of
$SO(2N)$. Again, using the maximal embedding of $SU(2)$ in $SU(N)$ and
corresponding representation of the basis Eq.(\ref{sungenerators}),
it remains to show that the generators are Hermitian and antisymmetric.
The symmetric tensor $C_{i_1 \cdots i_n}$ relevant for vector harmonics
satisfying the involution Eq.(\ref{involution}) allows only odd numbers of
$\Sigma_3$. It is now convenient to make $(\Sigma_1, \Sigma_2, \Sigma_3)_{D_2}
= (\Sigma_3, \Sigma_1, \Sigma_2)_{S_2}$. This cyclic permutation
of $N$-dimensional $SU(2)$ generators redefines $\Sigma_3$ naturally into an
anti-symmetric matrix :
\be
\Sigma_3 = {\rm Sym} [ \sigma_2 \otimes {\bf I} \otimes \cdots \otimes {\bf I}
+ {\bf I} \otimes \sigma_2 \otimes {\bf I} \otimes \cdots \otimes {\bf I}
+ {\bf I} \otimes {\bf I} \otimes \cdots \otimes \sigma_2].
\ee
Noting that the constant tensor $C_{ijk\cdots}$'s
are completely symmetric, we find that only a set of generators left over
are $2N(2N-1)/2$ independent, $2N \times 2N$ Hermitian, anti-symmetric
matrices. They are the generators of $SO(2N)$.
This is in precise agreement with the fact that heterotic M(atrix) theory is
described by SO(2N) gauge theory.

Again, in the large $N$ limit, the non-commutativity among the surviving
$T_{(i)}$'s die off sufficiently fast that the resulting $SO(N)$ Lie algebra
is exactly the same as the area-preserving diffeomorphism algebra
${\rm Diff}_0 (D_2)$.

One can find an exact solution of M(atrix) Q-balls of disk shape. Note that
the $\integer_2$ involution retains the ${\bf l} = 1$ sector of the spherical
harmonics and corresponding SU(2) generators. The magnetic quantum number is
all projected to ${\bf m} = 0$, but this is again compatible with the closure 
of ${\bf l} = 1$ sector.
Thus quasi-static, non-rotating disk Q-ball can be described by $\integer_2$
involution of the spherical Q-ball:
\be
{\bf X}^i (\tau) = \hbar X(\tau) \cdot (\Sigma^i \otimes \sigma_3 )
\ee
Similarly, the disk Q-ball rotating around the symmetry axis is described by 
\be
{\bf X}^i (\tau) = \hbar X(\tau) \cdot \Omega^{ij} \cdot
(\Sigma^j \otimes \sigma^3)
\ee
Both configurations carry `Chan-Paton' quantum number of the end-of-world
nine brane at which they are attached. As such, as the quasi-static
non-rotating disk Q-ball decays eventually, we expect that the energy is
radiated away not only via eleven-dimensional graviton and antisymmetric 
tensor field but also ten-dimensional gauge boson.

Next, consider real-projective Q-ball. This is described by an involution 
${\bf x} \rightarrow - {\bf x}$ of the sphere $S_2$. Under the involution 
the spherical harmonics maps as $Y_{\bf lm} \rightarrow (-)^{\bf l}
Y_{\bf lm}$. Hence, a complete set of vector
harmonics that generate the area-preserving diffeomorphism ${\rm Diff}_0
(RP_2)$ are the odd-parity subset of $S_2$ spherical harmonics
Eq.(\ref{sphericalharmonics}):
\be
L_{\bf lm} = \{  Y_1, \,\, Y_3 , \,\, Y_5 , \, \cdots, \, Y_{2N-1} \}
\ee
Hence, the Poisson algebra among these subset of harmonics is isomorphic
to sub-algebra that closes among the generators:
\be
T^{(1)}, \,\,T^{(3)}, \,\,T^{(5)}, \cdots, T^{(2N-1)}.
\label{subset}
\ee
This sub-algebra forms~\cite{poperomans} $Sp(2N, {\bf C}) \, \cap \, 
SU(2N) = USp(2N)$ group. To see this, consider a totally anti-symmetric 
$(2^{(2N-1)} \times 2^{(2N-1)})$ matrix ${\cal M}$:
\be
{\cal M} \equiv {\rm Sym} [ \sigma_2 \otimes \sigma_2 \otimes \cdots
\otimes \sigma_2].
\label{symplecticmetric}
\ee
It is straightforward to check that the $SU(2)$ generators $\Sigma_i$
in Eq.(\ref{su2generators}) of the $2N$-dimensional representation
satisfies:
\be
\Sigma_i \cdot {\cal M}  + {\cal M} \cdot \Sigma_i^{\rm T} = 0,
\ee
hence, for $i = 1, 3, \cdots, (2N-1)$,
\be
T_{(i)} \cdot {\cal M} + {\cal M} \cdot T_{(i)}^{\rm T} = 0.
\ee
Therefore, the subset of generators Eq.(\ref{subset}) forms an
$Sp(2N, {\bf C})$ algebra. Since they are Hermitian as well, the
generators actually closes under $Sp(2N, {\bf C}) \cap SU(2N) = USp(2N)$.

Exact solution of the real-projective M(atrix) Q-ball is obtained 
again by noting that the ${\bf l}=1$ spherical harmonics modes are closed
under the Poisson bracket. Thus, the Q-ball configuration is given by
\bee
{\bf X}^i(\tau) &=& \hbar X(\tau) \cdot (\Sigma^i \otimes \sigma^3)
\nonumber \\
&=& \hbar X(\tau) \cdot \Omega^{ij} (\tau) \cdot
(\Sigma^i \otimes \sigma^3)
\eee
for non-rotating quasi-static and rotating real-projective Q-balls 
respectively. They do not carry `Chan-Paton' factors of end-of-world
nine brane. As such, decay of real-projective Q-ball is qualitatively
the same as the spherical Q-ball except that the total energy is half of 
the latter.

The disk and real-projective space Q-balls and black holes formed thereof
are of some interest. First of all, the formed black hole will
entail non-orientable spacetime. Such black holes have been discussed in 
different context~\cite{gibbons}. Moreover, in heterotic M(atrix) theory, 
black holes that are formed out of disk are charged under the 
`twisted sector' gauge group. These gauge groups are localized at the 
end-of-world nine-brane.
Therefore, as the disk Q-ball radiates its energy, the branching fraction
that are emitted via gauge bosons  will propagate only in
ten dimensions, while the radiation via graviton and antisymmetric tensor 
fields will be emitted to the full eleven dimensions.  

%%%%%%%%%%%%%%%%%%%%%%%%%%%%%%%%%%%%%%%%%%%%%%%%%%%%%%%%%%%%%%%%%%%%%%%
\section{Discussions}
%%%%%%%%%%%%%%%%%%%%%%%%%%%%%%%%%%%%%%%%%%%%%%%%%%%%%%%%%%%%%%%%%%%%%%%
In this paper, we have studied M(atrix) Q-balls -- quasi-static membranes 
of spherical, disk or real-projective topology. 
We have studied both kinematics and dynamics of the M(atrix) Q-balls 
in detail. In particular, we have found that the long-range `monopole' field 
produced by the Q-balls are correctly described by the M(atrix) theory 
in the large N limit.

Q-balls of more complicated topology should be equally possible. 
For example, for genus one, there are four classes of M(atrix) membranes 
topology: torus, cylinder, M\"obius strip and Klein bottle. Again, 
instantaneous configuration of these Q-balls have been found explicitly 
in~\cite{kimrey}.
In uncompactified transverse space ${\bf R}_9$, all these Q-balls are 
unstable  and eventually undergo gravitational collapse. Long-distance 
property of these collapsed Q-balls should be dominated by the `monopole'
component of the Q-ball energy distribution. That is, we expect that 
long-range gravitational field is the same as the result 
obtained in section 3 for Q-balls of {\sl any} topologies in so far as the
total Q-ball energy is the same.
An interesting question is whether it is possible to distinguish the 
topology of Q-balls by probing carefully the potential generated by 
higher moments of the energy distribution. 
This will be reported elsewhere~\cite{toappear}.

Even more interesting but far more demanding question is regarding the final
fate of the Q-balls. Eventually, due to gravitational collapse, they will
all fall inside their Schwarzschild radius and form a black hole. 
Superficially, it appears that M(atrix) Q-balls cannot form a Schwarzschild
black hole since the Q-ball self-energy depends on its area for all dimensions
in contrast to the behavior of Schwarzschild black hole. However, in order to
analyze the final stage of black hole formation, one has to gain a better
understanding of possible Q-ball configuration in the large N limit. What
happens if the M(atrix) Q-ball shrinks? Eventually, the Q-ball size will become smaller than the M-theory Planck scale. At this stage, it should be possible
for the macroscopic Q-ball to self-intersect and self-interact strongly 
and disintegrate into N miniature Q-balls made out of just two D0-partons. 
These miniature Q-balls will then attract gravitationally one another and form
a new kind of Q-ball bound-state. Unlike large N Q-balls the miniature Q-balls
exhibit interesting dynamical cross-over to an ergodic system~\cite{chaos}. 
This novel possibility may then offer a resolution to the above puzzle. 
In parallel to recent works~\cite{susskind, klebanovsusskind, horowitzmartinec, susskind2}, it would be interesting to check the thermodynamic
relations of the black holes formed out of Q-balls via M(atrix) theory itself.

In addition, once formed, the Q-ball black holes will decay quantum mechanically via Hawking radiation. It 
would be extremely interesting to see if Hawking radiation can be systematically analyzed using the gravitationally collapsing Q-balls. Work in this 
direction is in progress. 

\vskip 1.0 cm
\centerline{\bf Acknowledgments}
I thank N. Kim, J.-T. Lee, Y.-S. Myung for useful discussions and C. Zachos
for earlier correspondences.

%%%%%%%%%%%%%%%%%%%%%%%%%%%%%%%%%%%%%%%%%%%%%%%%%%%%%%%%%%%%%%%%%%%

\end{document}